\documentclass{basi}
\usepackage{graphics}
\usepackage[british]{babel}
\usepackage[varg]{txfonts}
\usepackage{rotating}
\usepackage{dcolumn}

\begin{document}
\title[On the Triggering of M-Class Solar Flare due to Loop-loop Interaction
in AR NOAA 10875]{On the Triggering of M-Class Solar Flare due to Loop-loop Interaction
in AR NOAA 10875}
\author[Pankaj Kumar et~al.]%
       {Pankaj Kumar$^1$\thanks{email: \texttt{pankaj@kasi.re.kr}},
       Abhishek K. Srivastava$^{2}$, B.V. Somov$^3$, P.K. Manoharan$^4$, R. Erd\'elyi$^5$, Wahab Uddin$^2$\\
       $^1$Korea Astronomy and Space Science Institute (KASI), Daejeon, 305-348, Korea.\\
       $^2$Aryabhatta Research Institute of Observational Sciences (ARIES), Nainital-263129, India.\\
       $^3$Astronomical Institute, Moscow State University, Universitetskij Prospekt 13, Moscow 119992, Russia.\\
       $^4$Radio Astronomy Centre, NCRA, Tata Institute of Fundamental Research, Udhagamandalam (Ooty) 643 001, India.\\
       $^5$Solar Physics and Space Plasma Research Centre (SP2 RC), School of Mathematics \& Statistics, The University of
Sheffield,\\
 Sheffield S3 7RH, UK.
       }

\pubyear{2011}
\volume{00}
\pagerange{\pageref{firstpage}--\pageref{lastpage}}

\date{Received \today}

\maketitle
\label{firstpage}

\begin{abstract}
We present multiwavelength analysis of an M7.9
/1N solar flare which occurred on
27 April 2006 in AR NOAA 10875. The flare was triggered due to the interaction of two loop systems. GOES
soft X-ray and TRACE 195 \AA \ image sequences show the observational evidences of 3-D X-type loop-loop interaction with converging motion at the interaction site. 
We found the following characteristics during the loop-loop interaction: (i) a short duration/impulsive flare obeying the Neupart effect, (ii) double peak structure in radio flux profiles (in 4.9 and 8.8 GHz), (iii) quasi-periodic oscillations in the radio flux profiles for the duration of $\sim$3 min, (iv) absence of CME and type III radio burst.
The above characteristics observed during the flare are in agreement with the theory and simulation of current loop coalescence by Sakai et al. (1986). 
These are unique multiwavelength observations, which provide the evidences of loop-loop interaction and associated triggering of solar flare without CME.
\end{abstract}

\section{Introduction}
The reconnection between magnetic flux tube is important to explain the solar flare. The pioneering work in this direction was done by Gold and 
Hoyle (1960), who proposed a model of current loop interaction in solar flare. Later on, Sweet (1969) included in the modelling the importance of current sheet formed in between the region of interacting loops with antiparallel field directions. The process of loop coalescence may be different depending on the geometry of the interaction region (Sakai \& de Jager 1991). There are three possible configurations on the basis of loop radius (R) and length of interaction region (L), (i) if L$>>$R, 1-dimensional current sheet is induced in between the quasi-parallel loops (known as 1-D, I-type coalescence). (ii) L$>$R, 2-D Y-type coalescence. (iii) L$\approx$R, 3-D X-type coalescence. The triggering of solar flare needs an initial instability (i.e. footpoint rotation or shear motion), causing the approach of two or more flux tubes. However, YOHKOH Soft X-ray images have shown some observational evidences of above current-loop coalescence (Shimizu et al. 1992). High resolution multiwavelength observations can provide a better understanding about the processes/mechanism of current-loop coalescence. In this paper, we present a rare multiwavelength observations of 3-D X-type current-loop coalescence, which is in agreement with the earlier proposed theories and models of interacting loops proposed by Sakai et al. (1986). 
 
\begin{figure}
\centerline{
\includegraphics[width=8.2cm]{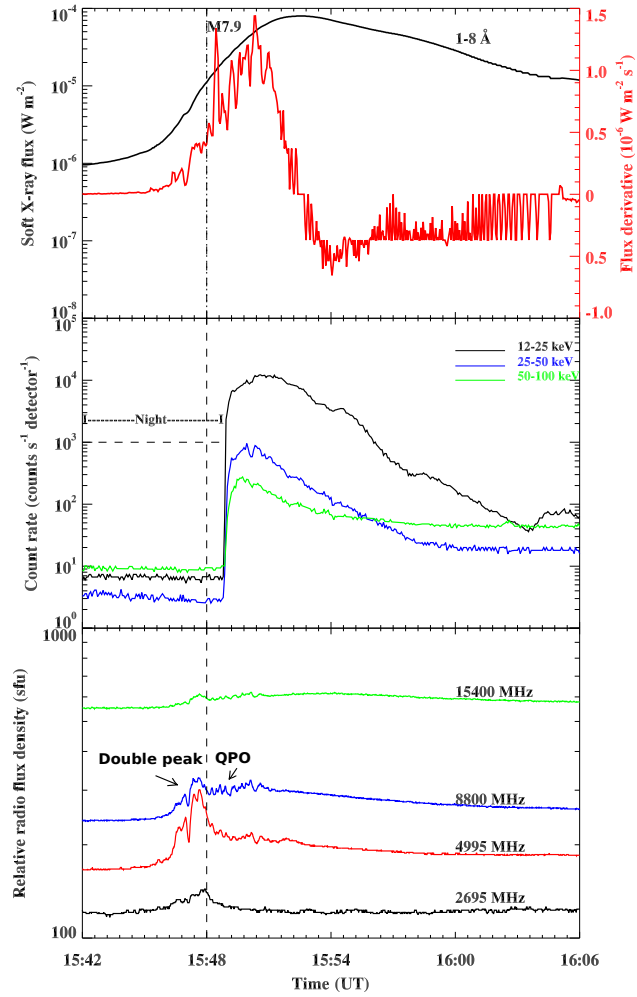}
}
\caption{GOES soft X-ray flux, soft X-ray flux derivative, RHESSI hard X-ray fluxes in three different energy bands and RSTN radio flux profiles on 27 April 2006. Radio flux profiles in 4.9 and 8.8 GHz show the double peak structure and quasi-periodic oscillations (QPO).}
\label{flux}
\end{figure}

\begin{figure}
\centerline{
\includegraphics[width=5cm]{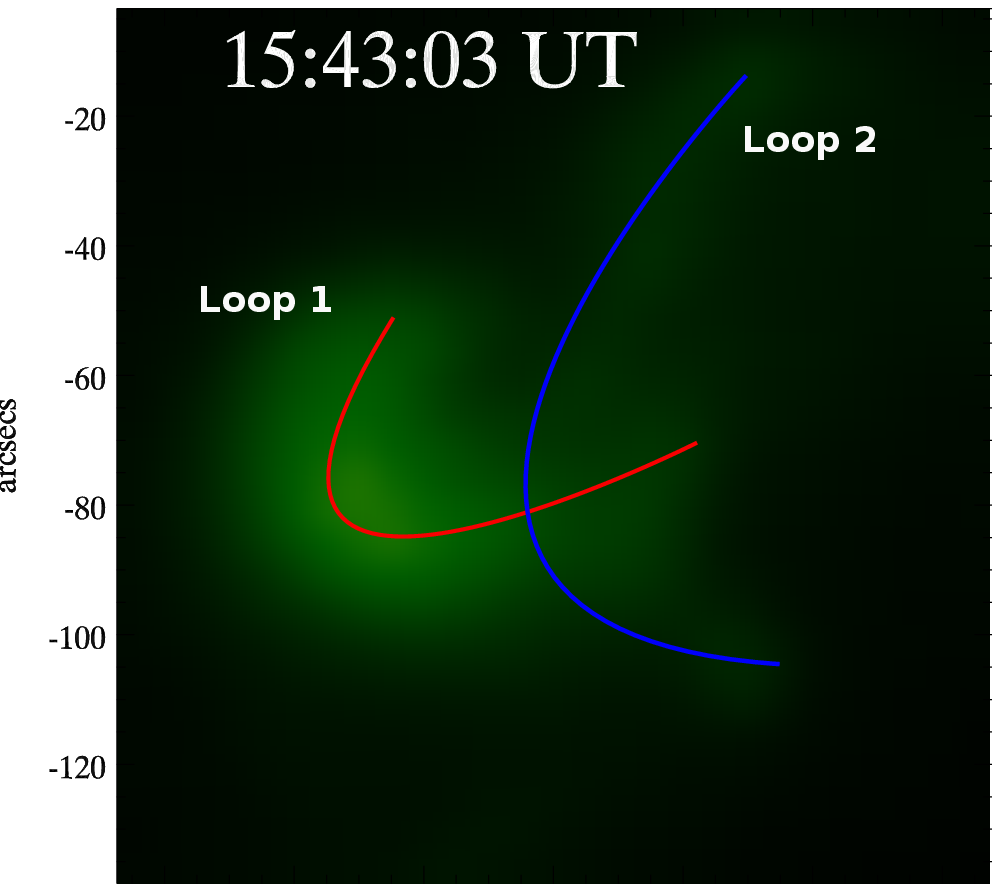}
\includegraphics[width=5cm]{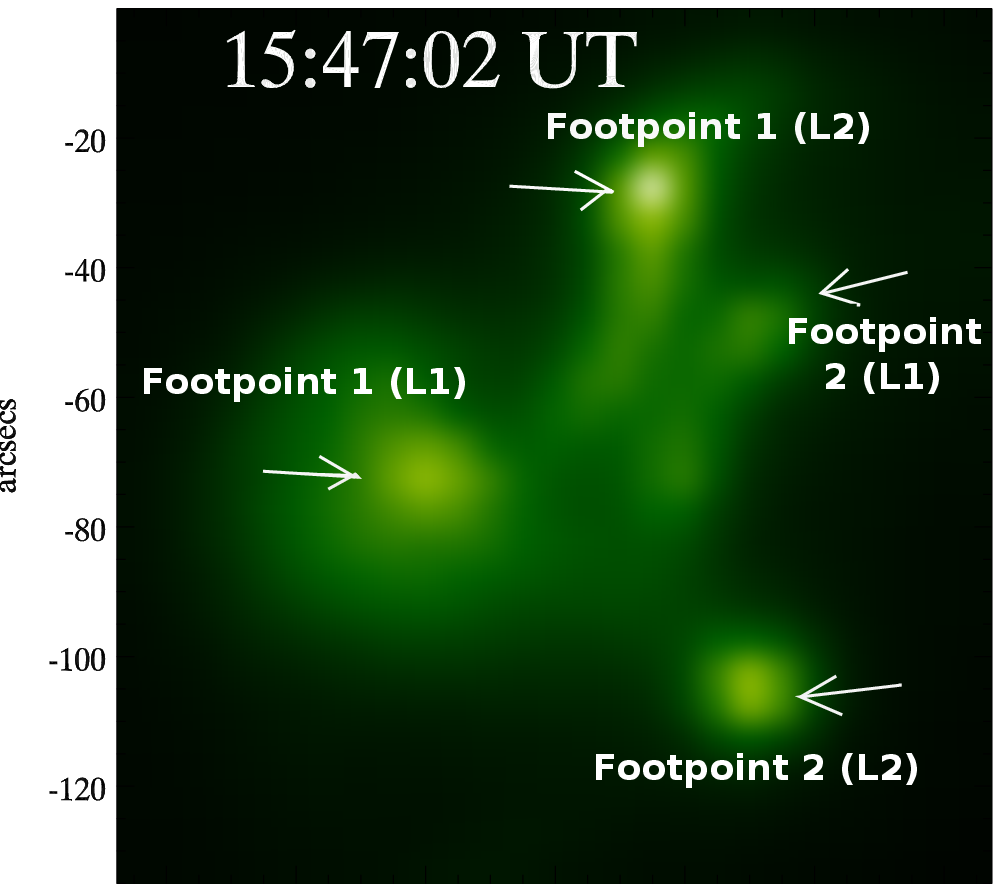}
}
\centerline{
\includegraphics[width=5cm]{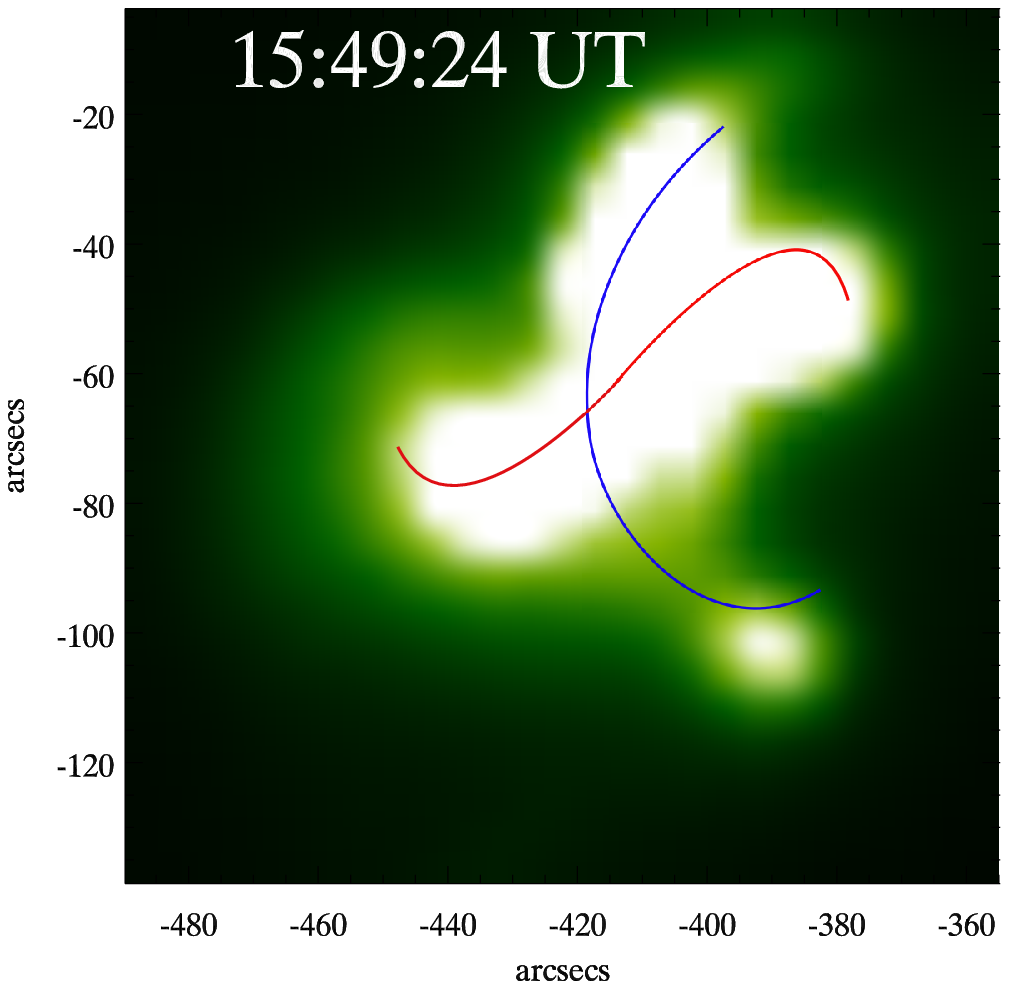}
\includegraphics[width=5cm]{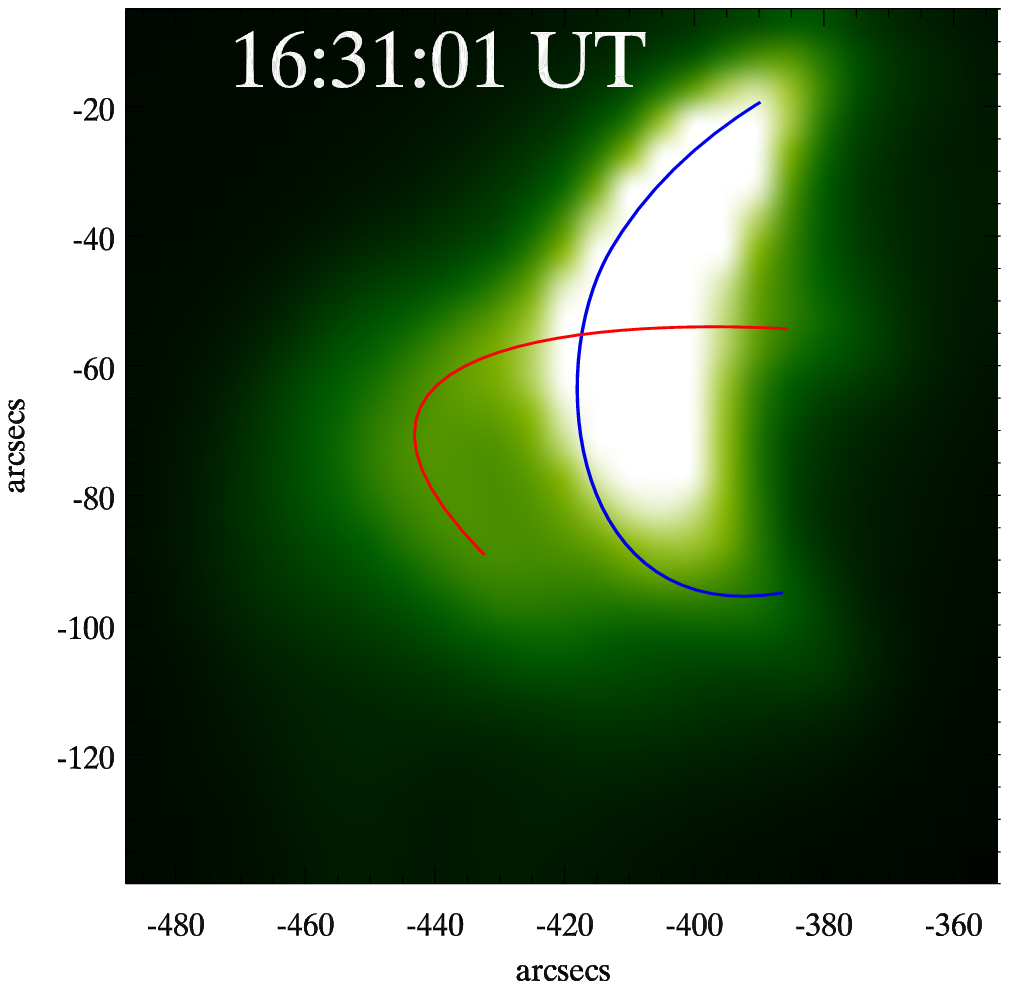}
}
\caption{GOES soft X-ray images (6-60 \AA) showing the interacting loops and associated flare on 27 April 2006.
Red and blue curves indicate the interacting loop 1 (L1) and loop 2 (L2) respectively.}
\label{sxi}
\end{figure}
\begin{figure}
\centerline{
\includegraphics[width=5cm]{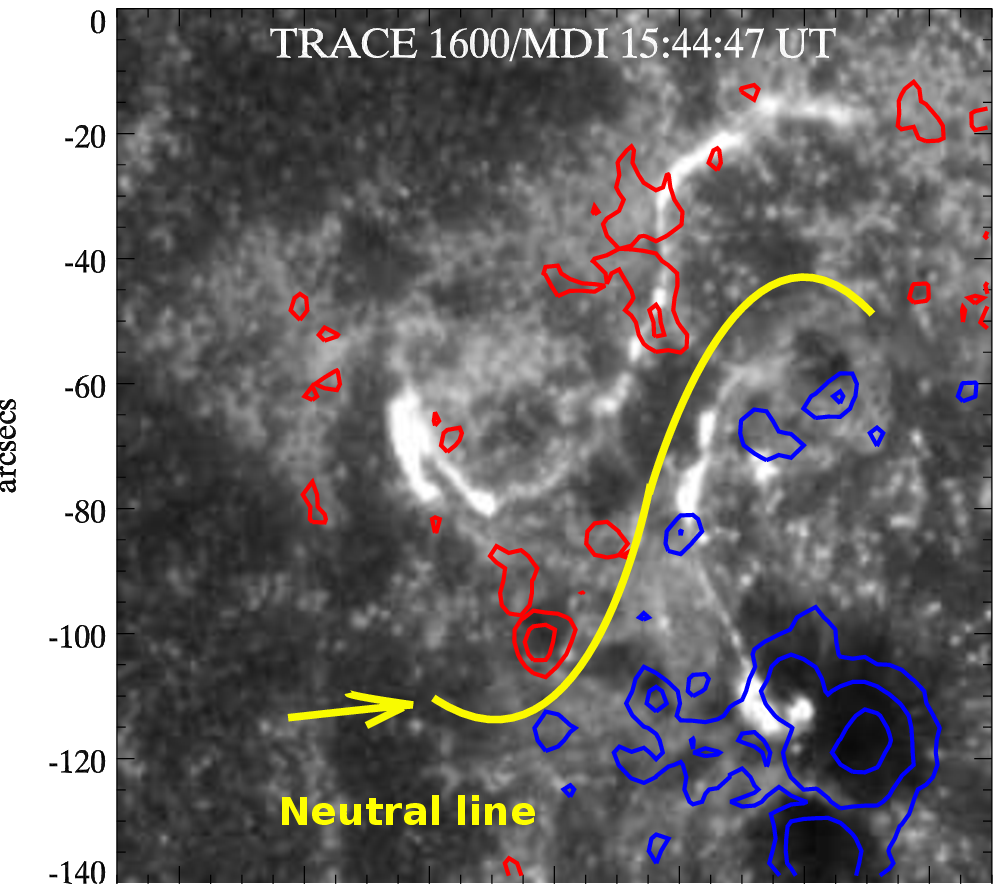}
\includegraphics[width=5cm]{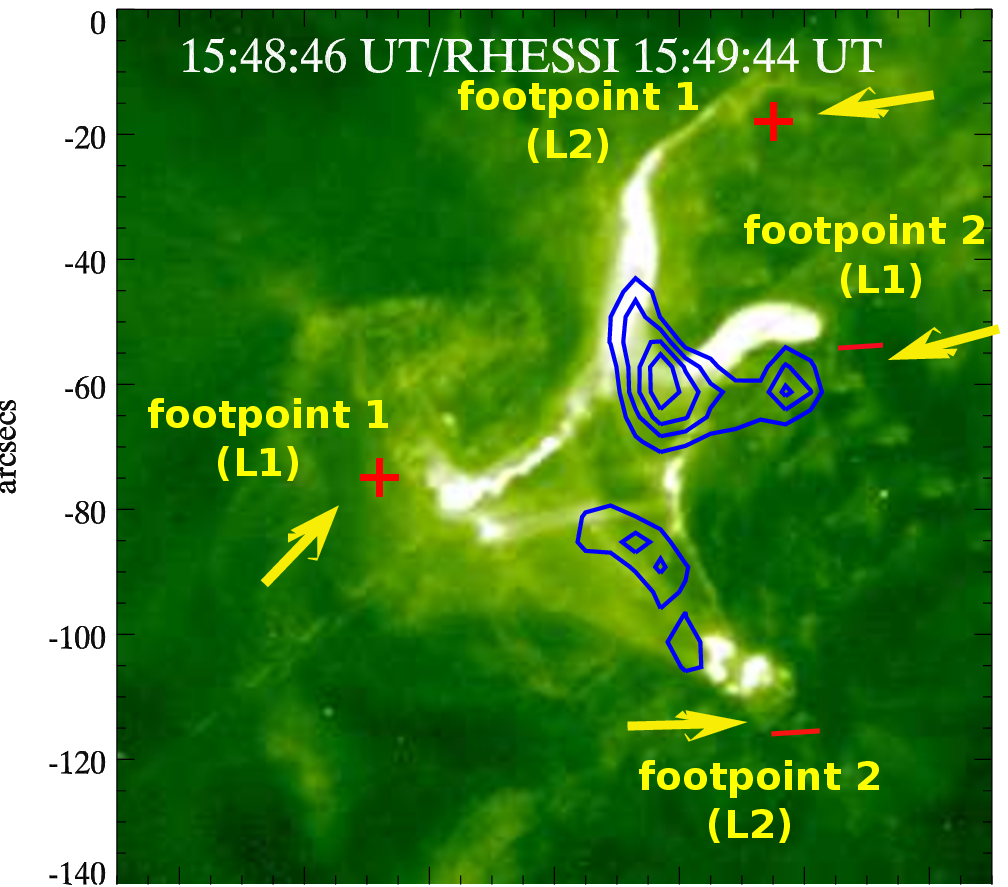}
}
\centerline{
\includegraphics[width=5cm]{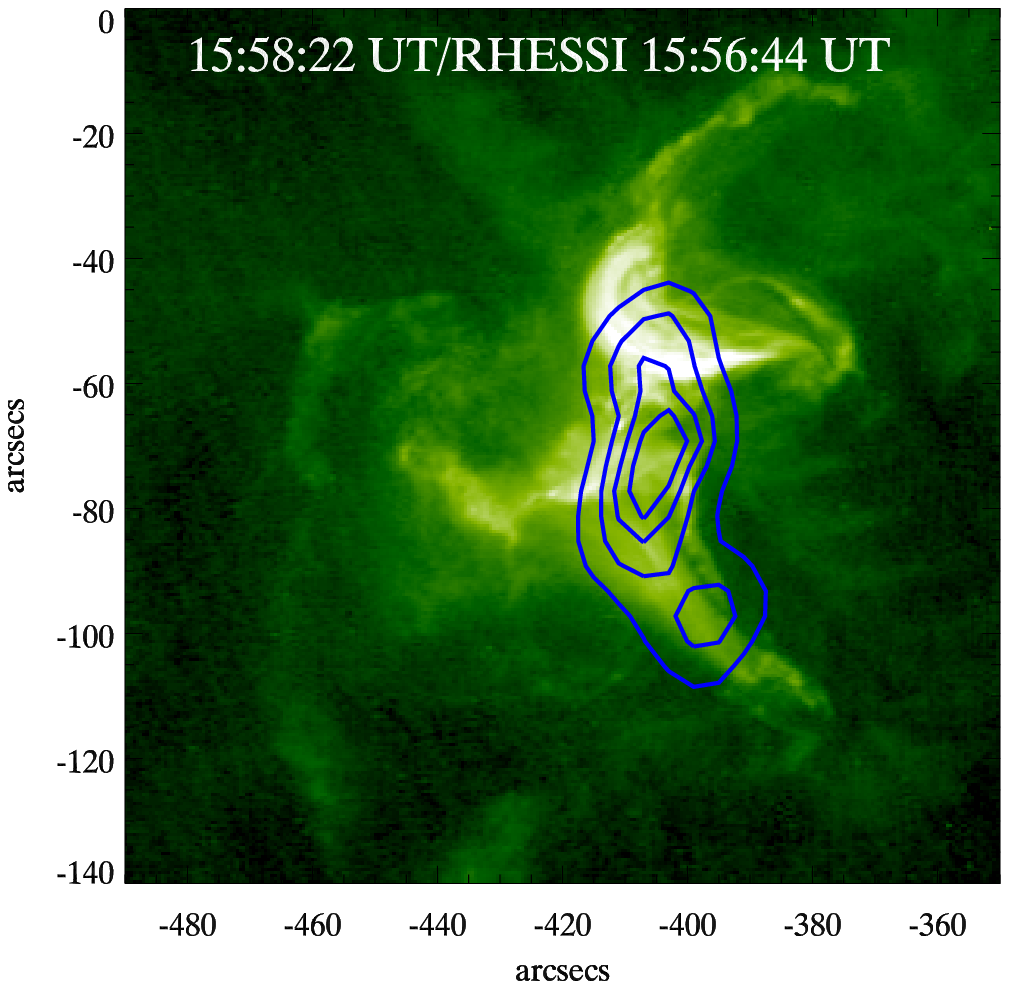}
\includegraphics[width=5cm]{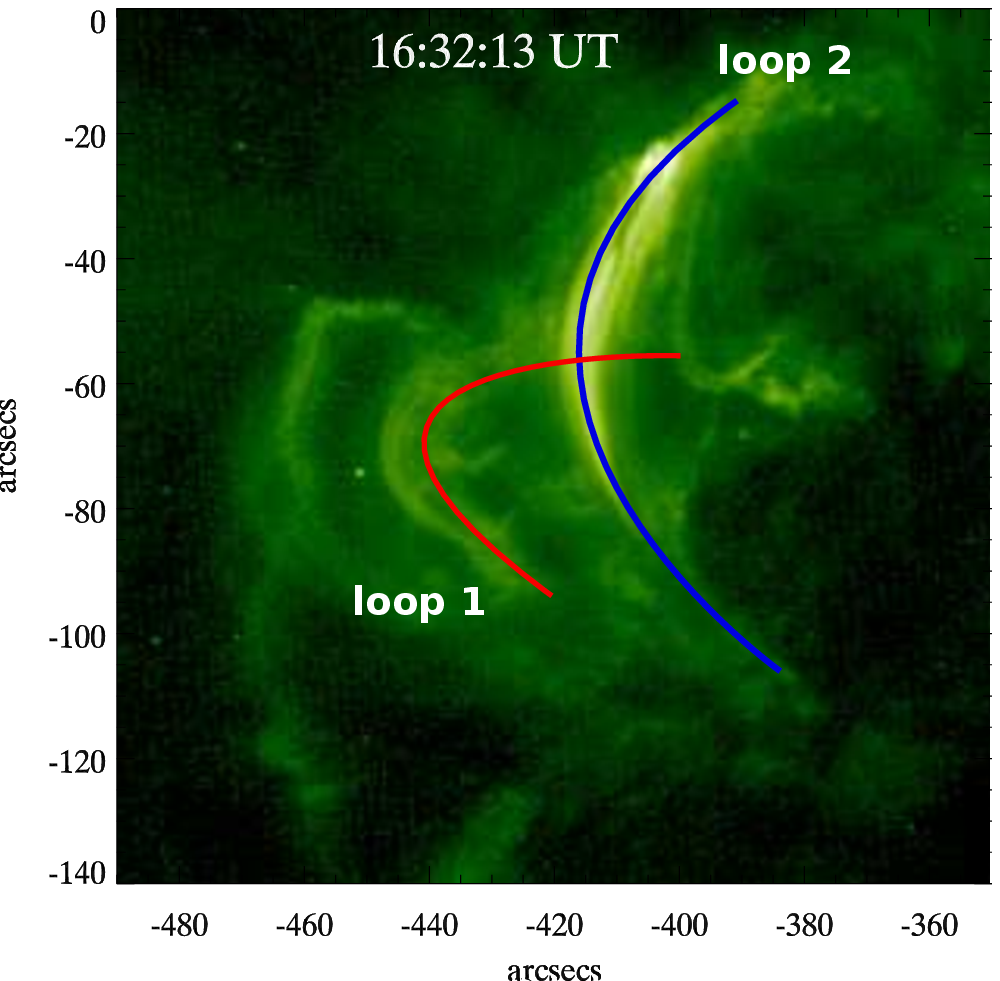}
}
\caption{(i) Top-left: TRACE 1600 \AA \ image overlaid by SOHO/MDI magnetogram contours of positive (red) and negative (blue) polarities. The contour levels are $\pm$500, $\pm$1000, $\pm$2000, $\pm$3000 G (gauss). Neutral line is drawn by green line. (ii) Top-right and bottom-left: TRACE 195 \AA \ EUV images overlaid by RHESSI hard X-ray contours (12-25 keV) showing the picture of interacting flux-tubes. RHESSI contour levels are 20$\%$, 40$\%$, 60$\%$, 80$\%$ of maximum intensity level. (iii) Bottom-right: TRACE 195 \AA \ image during the flare decay phase showing the relaxed loop-system drawn by red and blue lines.} 
\label{sxi}
\end{figure}
\begin{figure}
\centerline{
\includegraphics[width=4.4cm]{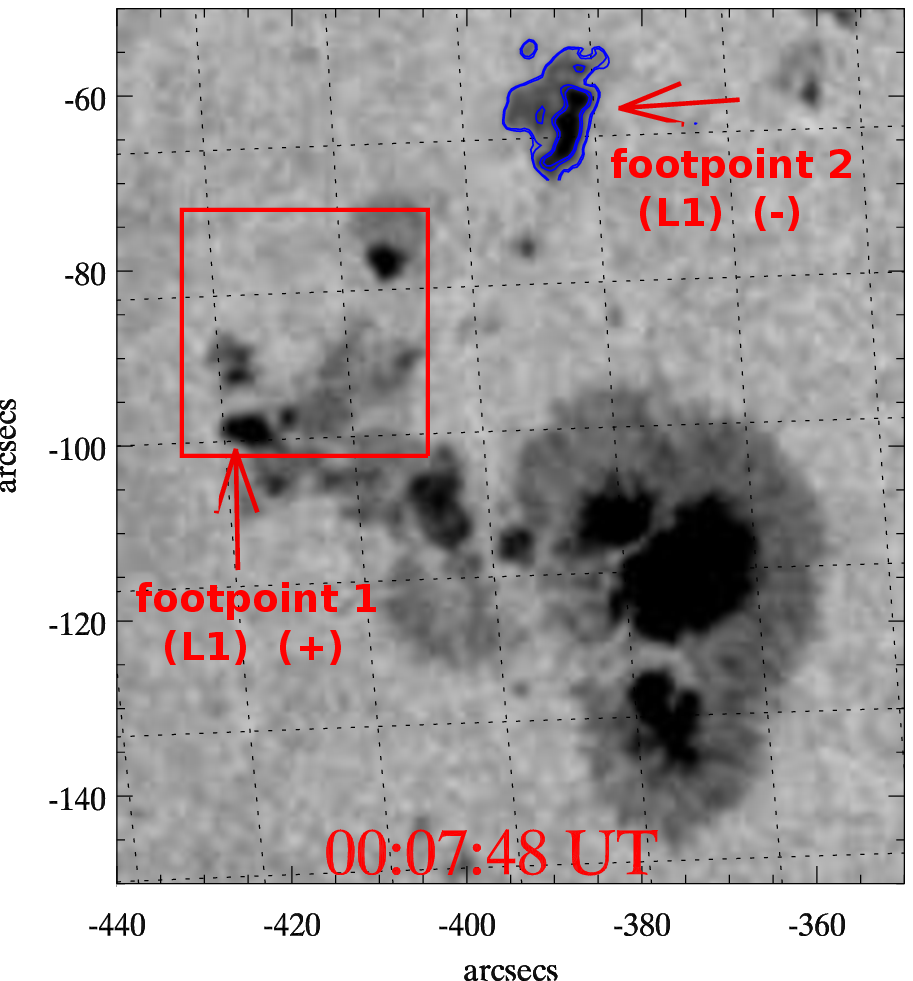}
\includegraphics[width=4cm]{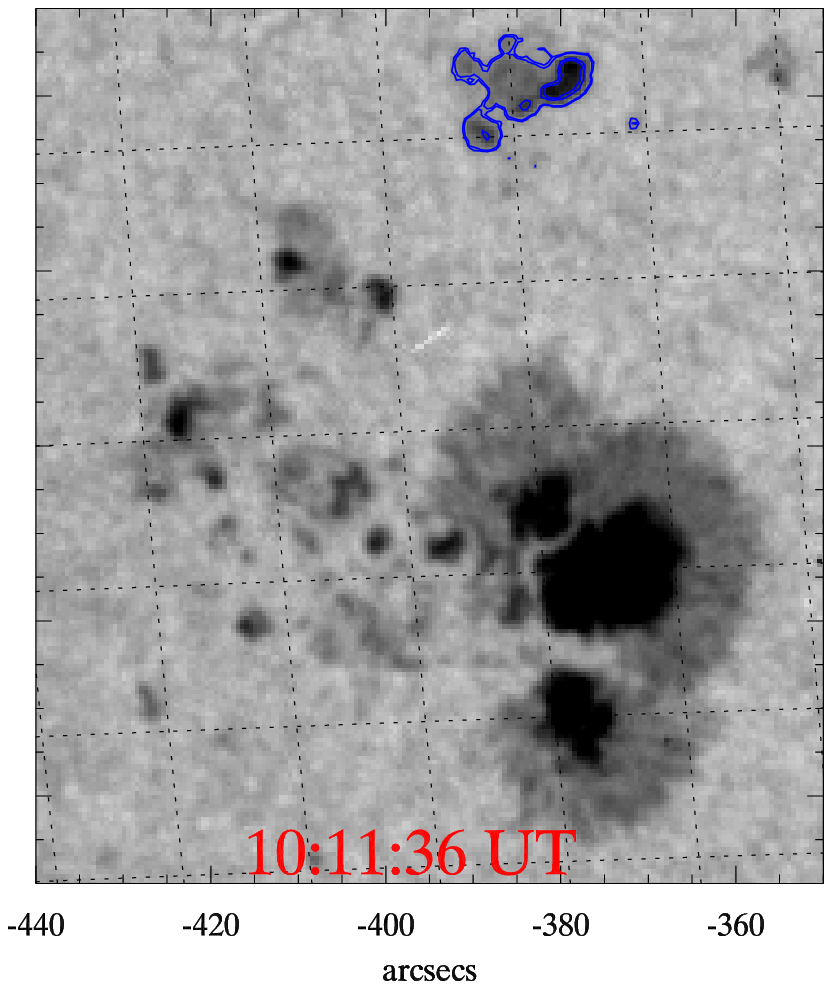}
\includegraphics[width=4cm]{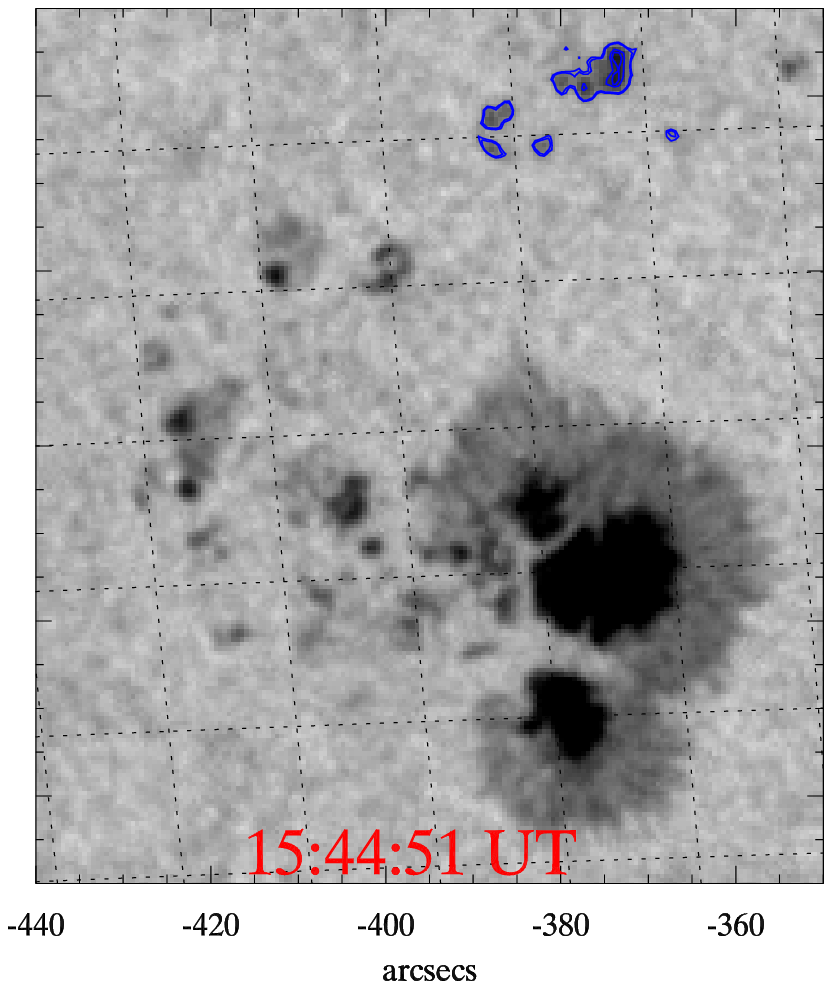}
}
\caption{TRACE White-light images of the active region NOAA 10875 on 27 April, 2006. Blue contours indicate the negative polarity spot corresponding to footpoint 2 of loop 1 (L1), which showed the shear motion before triggering the solar flare.}
\label{sxi}
\end{figure}
\section{Observations and Results}

Active Region NOAA 10875 was located at $\sim$S10E20 on 27 April 2006 with magnetic 
field configuration of $\beta\gamma\delta$. Figure 1 displays the 
flux profiles in GOES soft X-rays (top), RHESSI  hard X-rays (middle) and radio 
(bottom) frequencies. The top panel shows the soft X-ray flux profile (in 1-8 \AA) 
 along with its derivative (red). According to GOES soft X-ray flux profile,
 an M7.9 flare was started at $\sim$15:45 UT, peaked at $\sim$15:52 UT and ended at $\sim$15:58 
 UT. This was a short duration and impulsive flare. The middle-panel shows the hard X-ray flux profiles in 12-25 (black), 25-50 (blue) and 50-100 keV (red) energy bands. The soft X-ray flux derivative matches well 
 with the hard X-ray flux. It means that the accelerated electrons, which produces the hard X-ray 
 emission also heat-up the plasma and generate the soft X-ray emission. This flare satisfies the 
 Neupart effect. The bottom-panel shows the RSTN 1 sec cadence radio flux profiles in 2.6, 4.9, 
 8.8 and 15 GHz frequencies observed at Sagamore Hill. These profiles show significant variations 
 during the flare. The double peak structure and quasi-periodic oscillations (QPO) are evident in 4.9 and 
 8.8 GHz frequencies (indicated by arrows). 
 
 Figure 2 displays the selected images from GOES Soft X-ray Imager (SXI) in 6-60 \AA \ ($\sim$2 MK). It observes 
 512$\times$512 pixels images (5$^{\prime\prime}$ per pixel resolution) with $\sim$1 min cadence. The top-left image at 15:43 UT shows the coronal view of the active region site before the flare activity. This image shows the two loops (loop 1 and loop 2) oriented in different directions (indicated by red and blue curved lines). Before the flare activity, loop 1 seems to be more brighter in comparison to loop 2. Top-right image displays the image at 15:47 UT during the flare impulsive phase. Four bright foot-points of the associated loop systems are observed in this image (indicated by arrows for loop 1 and loop 2). The bottom-left image shows the picture of interacting bright loops before the flare maximum. Loop 1 shows some orientation change (`S' shape) during the interaction. This image clearly shows the crossing/touching loop systems as a result of `chromosphric evaporation' as predicted by Somov. The relaxed loops are shown in the bottom-right image during the decay phase of the flare. It should be noted that the orientation of loop 1 is changed during the flare activity, whereas loop 2 orientation is not changed. The loop 1 shows `S' shaped configuration during the flare, which is the signature of helicity in this loop. Therefore, loop 1 seems to be responsible for the flare triggering. Another important point is that four bright footpoints (i.e. kernels) during the impulsive phase of the flare are most likely generated during the nonthermal particle acceleration from the interaction region in the corona. Therefore, the soft X-ray images show a unique and rare view of interacting loop system. 
 
Figure 3 displays the TRACE EUV images in 1600~\AA \ and 195 \AA \ wavelength bands for the chromospheric and coronal views, respectively. The top-left image shows the TRACE 1600~\AA \ image overlaid by SOHO/MDI magnetogram contours. Red contours show the positive polarity, whereas blue ones show the negative polarity. Two bright ribbons are observed on the both sides of neutral line(indicated by green line) during the flare initiation. The top-right image displays the TRACE 195~\AA \ (Fe XII) image during the flare impulsive phase. It shows the interacting loops indicated by L1 and L2. Four footpoints of the interacting flux tubes can be identified in this image. The careful comparison of this image with the previous one reveals that the 2 footpoints of the interacting loops were anchored in negative polarities (right side of neutral line) whereas 2 footpoints are  anchored in the positive polarities (left side of neutral line). The footpoints polarity of both loops (L1 and L2) is marked by `+' and `-' symbols. The bottom-right image shows the relaxed loop-systems during the flare decay phase and is in agreement with SXI image at 16:31 UT. In addition, to see the evolution of hard X-ray sources, we have overlaid RHESSI hard X-ray sources contours (blue) over TRACE images (top-right and bottom left) during the flare and associated interaction of loop-systems. Initially, we see 2 loop-top sources (top-right) of 2 current loops and later these 2 sources are merged into a single loop-top source (bottom-left). The merging of 2 loop-top sources into single one also confirms the interaction of two loops.
The soft X-ray image of the flare clearly reveals the two large solar loops
crossing to each other and exhibit the X-type interaction. We illustrate these features of the interacting
loop-systems in terms of the topological models (see ch. 3 in Somov (2007)). The radio flux profiles in 4.9 and 8.8 GHz exhibit double peak structures
 associated with interacting loop-systems (see bottom panel of Figure 1). The second double peak is stronger in comparison to the first one, which shows that
the more superthermal electrons accelerated from a higher amount of pre-accelerated electrons
generated the last double peak (Karlick{\'y} \& Ji{\v r}i{\v c}ka 2003). After this burst, we observe quasi-periodic oscillations specially in 4.9, 8.8 and 15 GHz frequencies
for the duration of $\sim$3 minutes, which may be attributed to modulations
by MHD oscillations or nonlinear relaxational oscillations of wave particle interactions.
Therefore, MHD waves can modulate the emissions from the trapped electrons (Aschwanden
2004).
We analyzed the TRACE White-light (WL) images to investigate the cause of loop coalescence. For this purpose, we check the footpoint motion of the corresponding loop systems. Figure 4 displays the selected TRACE WL images of the active region NOAA 10875 on 27 April, 2006. Blue contours indicate the negative polarity spot corresponding to footpoint 2 of loop 1 (L1), which showed the shear motion before triggering the solar flare. Therefore, the footpoint shear motion helped in the build-up of magnetic energy and generate instability in the lower-loop system, which collides with the higher loop system to trigger the solar flare.

\section{Discussion}
We have presented the rare multiwavelength obervations of loop-loop interaction and associated triggering of M-class flare. These are unique observations which show 3-D X-type coalescence of loop system. Moreover, especially from the radio signature the observation shows the double peak structure. Sakai et al. (1986) presented the physical characteristics of the explosive coalescence of current loops through computer simulation and theory and mentioned canonical characteristics of the explosive coalescence as (i) impulsive increase of kinetic energy of electrons and ions (ii) simultaneous heating and acceleration of particles in high and low energy spectra (i.e. Neupert effect) (iii) quasi-periodic amplitude oscillations in field and particle quantities (iv) a double peak (or triple peak) structure in these profiles. These observations clearly matches with all the above mentioned characteristics of the explosive coalescence. These observations confirms the theoretical predictions of loop-loop interaction. An initial instability is required for the approach of loop-systems. In this case, the shear motion of footpoint 2 (right, negative polarity) of loop 1 (red) helped in generating the instability, forming `S' shaped and colliding with the higher loop system (blue) and trigger the solar flare (Kumar et al. 2010). The absence of type III burst\footnote{http://urap.gsfc.nasa.gov/cgi/waves$\_$show$\_$png.py?yymmdd=20060427$\&$datatype=RAD2} during flare energy release confirms no opening of field lines. Moreover, no CME was observed during the flare event. Only the connectivity change took place during loop-loop interaction. In addition, merging of 2 hard X-ray loop-top sources into a single source also confirms the loop-loop interaction. Such type of events should be investigated/analyzed using high resolution data from SDO to shed more light on the mechanism and consequences of loop-loop interaction.

\section*{Acknowledgements}
We acknowledge GOES, TRACE, RHESSI, RSTN observations used in this study. PK thanks to KASI for providing the travel support to attend the conference. R.E. acknowledges M. K\'eray for patient encouragement and is also
grateful to NSF, Hungary (OTKA, Ref. No. K83133) for financial support received. This work is partially supported by CAWSES-India Phase II program, which is sponsored by ISRO.


\end{document}